\documentclass[pra,superscriptaddress,reprint,longbibliography, floatfix]{revtex4-1}
\usepackage{graphicx}
\usepackage{url}

\begin{document}

\title{Do female physics students benefit from informal physics programs they facilitate?}




 \author{Jessi Randolph}
  \affiliation{Department of Physics \& Astronomy, Texas A\&M University, College Station, Texas 77843}
  \email[send correspondence to: ]{etanya@tamu.edu}
  \author{Jonathan Perry}
  \affiliation{Department of Physics, University of Texas at Austin, Austin, Texas, 78712}
  \author{Jonan Phillip Donaldson}
  \affiliation{Center for Teaching Excellence, Texas A\&M University, College Station, Texas, 77843}
  \author{Callie Rethman}
  \affiliation{Department of Physics \& Astronomy, Texas A\&M University, College Station, Texas 77843}
  \author{Tatiana Erukhimova}
  \affiliation{Department of Physics \& Astronomy, Texas A\&M University, College Station, Texas 77843}

\date{\today}

\begin{abstract}
Gender bias, reduced sense of belonging, and lower physics self-efficacy are among the challenges faced by female students who choose to study physics. Prior studies focusing on this underrepresented group have examined the experiences and impacts of formal educational settings, leaving the impact of informal physics programs as a relatively overlooked area. Existing research on the impact of informal physics programs indicates that student facilitators, who help run the programs, can experience positive impacts on their learning and sense of community beyond the formal setting of a classroom or laboratory. In this study we took a first step, narrowing our focus to explore the relation between facilitation of informal physics programs and female students' physics identity, persistence, mindset, and worldview. We analyzed survey responses (32) and interviews (11) collected from undergraduate and graduate female students at a large, land-grant university. Our results, based on self-reported data, showed a statistically significant shift in confidence of choice of major after facilitating informal physics programs. Analysis of interviews indicated a positive effect of facilitation of informal programs on female student interest and motivation with regards to the field of physics, improved performance and competence beliefs, and the development of characteristics indicative of a growth mindset. A semantic network analysis showed statistically significant interdependencies between positive persistence and constructs including growth mindset, important early undergraduate experiences, gender stereotype threat, external recognition, and confidence. 

\end{abstract}

\maketitle

\section{Introduction}
Gender imbalance between female and male students in US physics departments has been a long-recognized problem \cite{NSFPaper, Sax, McCullough, whitten2003works, porter2019women}. A vast majority of prior studies seeking to improve female student representation and retention in physics concentrated on student experiences in formal physics settings such as classrooms and labs. In this study, we focus on the impact of {\it informal} physics programs on female university students who facilitate these programs with a goal to better understand and potentially improve the experience of female students in physics. 

Extensive prior research has explored the factors contributing to gender disparities and documented efforts on recruitment and retention of female students in physics. Gender stereotypes, discrimination, feeling as outsiders in the field, and lack of recognition by peers are among the factors responsible for “why so few” females choose and stay in physics \cite{blue2018gender, skibba2019women, hodapp2015women, hill2010so, barthelemy2016, kalender2019female, avraamidou2021identities, Lewis2016, moshfeghyeganeh2021effect}. Gendered differences in physics self-efficacy and physics identity are additional factors that undoubtedly complicate the problem of underrepresentation of female students in physics \cite{Blue2019, kalender2020damage, Eddy2016, Sawtelle2012, Hazari2010, Kalender2019, nissen2016gender, lindstrom2011self, li2021effect, Perez2014, Hyater-Adams2018, Hazari2013a}. Female students often report low physics self-identity (they don’t see themselves as a physics person) which is known to correlate with their choice of being in physics related majors \cite{Hazari2010, Kalender2019, Hazari2013b, seyranian2018longitudinal}. 

The societal biases and gender stereotypes often keep female students who participated in physics in high school from pursuing a physics major as the next step of their career \cite{skibba2019women, hodapp2015women, hill2010so, Hazari2010}. The gender stereotype threat, or activation of a negative gender stereotype, can, in turn, impact female student performance in physics classes both in high school \cite{marchand2013stereotype} and at the university level \cite{Maries, miyake2010reducing}. Female students, who enter universities to major in physics or other STEM fields that require taking introductory physics classes, often find themselves a minority in their physics classes, which could aggravate gender stereotype threat and cultural biases that so many female students experience in their physics careers \cite{blue2018gender}. Gender stereotypes often lead females to believe that they don’t possess the innate brilliance and talent that are often believed to be required to study physics \cite{leslie2015expectations, Bian, dweck2013self}, or that their intelligence in the field of physics is fixed (fixed mindset) which makes them unfit for the field. A study by Marshman and colleagues found that female students who took introductory physics classes not only had a more “fixed” mindset compared to male students at the beginning of the course, but they developed an even “more fixed” mindset after the course completion \cite{marshman2018longitudinal}. 

Gender stereotypes could make females feel that they don’t belong to the field of physics which can be linked to attrition of female students \cite{Lewis2016, Hewitt, london2011influences, lewis2017fitting}. This attrition often occurs during the first two years of college when many female students leave physical science and engineering \cite{goodman2002final}. According to a review paper by Lewis and colleagues, on average, women are more likely to leave physics because they do not feel that they fit in and are accepted by their peers. Their findings showed that feelings of belonging could increase motivation and performance; higher sense of belonging was linked to higher persistence: undergraduate and graduate female students who reported higher sense of belonging to the field also reported stronger intentions to persist within the field of physics \cite{Lewis2016, lewis2017fitting}. 

Prior literature shows that other major factors helping female students to persist and be successful in the field are their interest and their domain-specific (physics) self-efficacy and perceived recognition \cite{Hazari2010, nissen2016gender, lindstrom2011self, hazari2017importance, li2021effect, marshman2018longitudinal, marshman2018female, verdin2021power, espinosa2019reducing, henderson2020gender}. Multiple studies have shown that female students often report low physics self-efficacy, which can be damaging to their performance and retention \cite{kalender2020damage, Sawtelle2012, lindstrom2011self, marshman2018longitudinal}. The atmosphere in a physics classroom with biased gendered beliefs, lack of recognition, and isolation can further damage females’ self-efficacy and their interest in physics. Studies of motivational beliefs of students taking introductory physics classes indicated that the gendered gap in self-efficacy and reduction of interest in physics kept increasing from the beginning to the end of the course \cite{marshman2018longitudinal, marshman2018female, li2021effect}.

Several studies showed that undeveloped physics self-identity and related low physics self-efficacy can be improved by perceived recognition by the role models in the field, such as students’ physics high school teachers or their university physics instructors. High expectations from instructors and recognition of accomplishments can help female students to be successful and persist in physics \cite{kalender2019female, kalender2020damage, Kalender2019, hazari2017importance}. Although the consensus is that all of the aforementioned constructs are important for female students, Hazari and colleagues warn that each of these constructs needs to be considered in contextualized studies which are limited in terms of generalizability \cite{hazari2020context}. 

The majority of the studies mentioned above, as well as decades of research comparing the performance between male and female students on standardized physics concept tests and exams \cite{miyake2010reducing, madsen2013gender, pollock2007reducing, kost2009characterizing, traxler2018gender, dew2021gendered, ozmetin2021}, were concentrated on student experiences in formal physics settings such as classrooms and labs. In the current study, we focus on the experience of female students facilitating informal physics programs, often called physics outreach programs. In the presentation of our results and the discussion thereof, we will avoid, where possible, comparisons between male and female students in favor of simply focusing on the experiences of female students related to their participation in informal physics programs. 

There is an increased interest in understanding how student participation in informal physics programs provides a platform for the development of physics identity and aids in building a sense of community \cite{Hinko2016, Hinko2012, Fracchiolla2020, rethman2021impact, Prefontaine2018, randolph2021, perry2021}, which are constructs potentially critical for female students’ success, retention, and persistence in physics. Informal physics programs are less structured and vary in frequency, scale, or target audience scale \cite{Fracchiolla2018}. They can be implemented by any physics department without significant changes in the curriculum. The facilitation of informal physics programs can enhance the educational experience of university students by providing them with rich teaching opportunities: students need to explain the concepts to audiences of different ages who are unfamiliar with physics. The exciting and often less structured environment of informal physics programs generates facilitators' ownership and enthusiasm of being ambassadors for science \cite{Hinko2012, Hinko2016, rethman2021impact, bennett2020refining}.

Hinko and Finkelstein emphasized the benefits of university student engagement in physics outreach programs suggesting that we consider the interactions between the universities and the public as partnerships rather than “outreach” by the universities to the public \cite{Hinko2012}. They studied a program at the University of Colorado Boulder called PISEC (Partnerships for Informal Science Education in the Community) and found a positive impact on university student attitudes towards teaching and learning as a result of facilitation of this program \cite{Hinko2012}. Fracchiolla and colleagues found that facilitation of informal physics programs had a positive effect on development of a university student’s discipline identity \cite{Fracchiolla2020}. Facilitation of informal physics programs can provide university students with opportunities for experiential learning, leadership and teamwork, peer mentoring and peer learning, communication skill development, and networking opportunities \cite{rethman2021impact}, skills needed for the 21st century careers \cite{Heron2016}. 

In our recent work, we examined the impact of different informal physics programs on a large number of graduate and undergraduate students facilitating these programs \cite{rethman2021impact}. Our mixed-methods study included five informal physics programs run by the Department of Physics \& Astronomy at Texas A\&M University. The programs differed in scale and frequency spanning from a large annual physics festival with thousands in attendance to year-round smaller-scale events. These findings, based on self-reported data from surveys and interviews, showed that facilitation of physics informal programs had a positive impact on students’ physics identity and sense of belonging to the physics community; students reported improvement of their communication, teamwork and networking, and design skills. Although female students were not a focus of our analysis, we observed that female students self-reported a positive link between participation in the physics informal programs and their sense of belonging, a parameter that could be crucial for female student retention in the field \cite{Lewis2016}.

To the best of our knowledge, there have been no published studies on the impact of informal physics programs on university female students specifically. This paper is the first in-depth step addressing this important question. Using an existing database \cite{rethman2021impact}, we examined how such important factors as the formation of female student physics identity, their important experiences, and persistence in the field were influenced by their participation in the informal physics programs. We hypothesized that female facilitators of informal physics programs would experience a positive impact on the development of their physics identity (as a result of increased interest and motivation, improved performance and competence beliefs, internal and external recognition, confidence and self-efficacy), growth mindset, and persistence in the field of physics. The findings of this study could provide a step forward for any physics department in addressing an undoubtedly complicated and multifaceted problem of gender imbalance among physics and physics-related majors.

\section{Framework \& Methods}
To explore the impact of facilitating informal physics programs on female students, we analyzed a subset of data collected from a previous mixed-methods study \cite{rethman2021impact}. This prior work collected 117 survey responses and conducted 35 interviews with undergraduate and graduate students who facilitated at least one informal physics program between 2013-2019. The survey consisted of questions targeting dimensions related to student physics identity as defined below, sense of belonging to the physics community \cite{Lewis2016}, and 21$^{st}$ century career skills \cite{Heron2016}. The survey incorporated a subset of items motivated by categories from the Colorado Learning Attitudes about Science Survey \cite{adams2006new} and additional items reflecting the goals of this project \cite{rethman2021impact}. The survey was distributed via email in fall 2019. Interviews were conducted with a volunteer pool of respondents from the survey. Each interview was conducted by a researcher who was unfamiliar with each interviewee. Didactic interview questions were developed in collaboration with learning scientists and probed for more in-depth experiences from facilitating informal physics programs. To provide a structure to interpret interview responses, the researchers looked to prior literature to develop a framework including constructs important to the female student experience in physics. This framework included dimensions of physics identity, persistence, and mindset.

Physics identity was defined as in Hazari et al. \cite{Hazari2010}, which included dimensions of belief in one's ability to understand physics content, recognition by self and others as being good at physics, and interest in the field as demonstrated by the desire to understand physics. This framework was expanded upon by including elements of the Dynamic Systems Model of Role Identity (DSMRI) which characterizes identity as context-specific self-perceptions, values, goals, emotions, and beliefs \cite{kaplan2017complex}. In addition, we defined learning and how learning occurs through the frameworks of Situated Learning Theory and Transformative Learning Theory. These frameworks incorporate assumptions, beliefs, perspectives, and habits of mind \cite{Mezirow2009, Kegan2009, LaveWenger1991}. Following Dweck's definition \cite{dweck2013self}, student mindsets were either defined as being \textit{fixed} if they discussed intelligence as being something that cannot be changed or \textit{growth} if they indicated that intelligence level can be changed. Persistence was framed to be either positive, where there was a demonstrated desire to continue in a physics degree or in the field in general, or negative, where there was a demonstrated desire not to continue with a physics degree or in the field. Related to persistence the framework included a dimension to record important student experiences which changed, or contributed to, their trajectory towards physics \cite{hazari2017importance}. Such important events may overlap with recognition prior to an individual being a physics student and have been seen to relate to a students' resilience which can be a major force for persistence in the major \cite{avraamidou2021identities}. 

Through the constructs mentioned above, a total of 18 codes were used to categorize statements from student interviews. These codes were organized into categories of (i) \textit{physics identity} (e.g. interest and motivation, recognition, performance and competence, confidence and physics self-efficacy), (ii) \textit{persistence} (positive or negative), (iii) \textit{mindset} (growth or fixed), (iv) \textit{worldview} (cooperative or competitive), (v) \textit{important experiences}, (vi) \textit{accountability} (to scientific community, leadership, individual roles), and (vii) \textit{gender stereotype threat}. Categories i, ii, iv, and vii are grounded in DSMRI, categories ii, vi are grounded in situated learning theory, and categories iii, v are grounded in transformative learning theory.

To code interviews, a team of four researchers met regularly to review and discuss the code book. A set of three interviews were coded, and then the team met to make revisions to the code book and to resolve differences in the coding process. Two researchers coded individually while the remaining two researchers coded as a team. All four researchers then coded all eleven interviews, meeting periodically for further discussions. After completing this process, we achieved an intercoder agreement of $\kappa>0.9$. We should note here that one of the authors is the founder and organizer of several programs in which students in this study participated; three other authors are former participants in multiple programs. 

To examine the impacts and connections among the codes present in this study, we looked at both the frequencies of individual codes, as well as the relationships between them using a semantic network analysis. Semantic network analysis uses social network analysis tools to analyze relationships within networks of ideas or, in the case of this study, codes \cite{DonaldsonAllenHandy2019}. This analysis begins with examining the likelihood that one idea or coded segment of the text appears near another idea. Then Pearson's correlations of code or idea co-occurrences were determined for each pair of codes. This produces a correlation matrices at the $p<0.05$ level. This correlation matrix is then used as input in the UCINET network analysis software \cite{borgatti2002ucinet} as 1-mode networks \cite{scott2011sage}. Centrality for each code is then calculated in relation to other codes in terms of distance or steps between one code and all other codes to produce Eigenvector centrality measures \cite{kadushin2012understanding}. These measures were then visualized using the NetDraw software in which each code is positioned with direct ties to other codes with which it was statistically significantly correlated, as well as indirect ties in the form of connections with other codes which are not significantly correlated with the first code, but have connections in the form of ties with other codes which are statistically significantly correlated with both \cite{valente2010social}. To illustrate what is meant by the preceding sentence, consider a situation where code A is correlated with code B, and code B is correlated with code C, but code C is not correlated with code A. Code A is then said to have an indirect connection to code C. The maps shown later in this paper are the result of the analysis described here. The color of the blocks is determined from a Girvan-Newman cluster analysis \cite{RN2901}. Larger nodes and a higher number of links correspond to the frequency and centrality of a code \cite{Scott2017, Kadushin2012}.

\section{Results}
In this section, we present results from a survey and interviews which were collected as part of a prior study \cite{rethman2021impact}. Our database of 117 completed surveys contained 32 responses from female students, including 20 undergraduate, 11 graduate, and one who was both. The majority of female respondents were physics majors, with the rest of the responses coming from engineering and other science majors. Interviews were conducted with 35 students, 11 of which were female students. Of the female interviewees, 6 were undergraduate students and 5 were graduate students. During interviews, students self-reported the impacts that their participation in informal physics programs had on their physics identity, values, abilities, and perceptions. Throughout this and the following sections, we will use the terms ``informal'' and ``outreach'' interchangeably to describe programs in which students participate that are outside formal curriculum. 

First, we analyzed the results from the survey. Self-reported data showed that male and female students had no statistically significant differences in their perceptions of the impact of informal physics programs on themes such as networking within the department, confidence in communication, and sense of belonging within the field. One theme which showed a statistically significant difference was confidence in choice of major prior to participating in informal physics programs. Using a Mann-Whitney U test \cite{ResearchMethodsInEducation}, we found that female students were less confident at the $p=0.01$ level, with a medium effect size, using Cohen's d with a Hedges' correction, of $d=0.75$. This difference disappeared after participating in informal physics programs, $p=0.49$. 

\begin{figure}
  \includegraphics[width=0.9\linewidth]{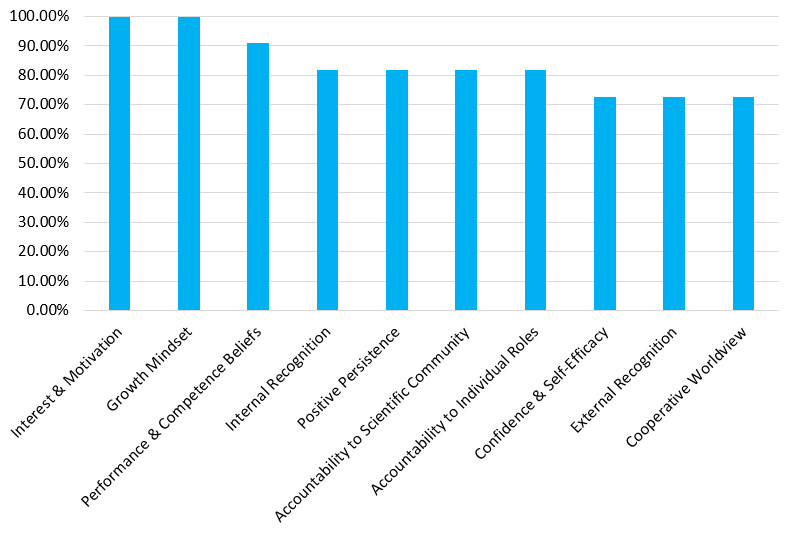}
  \caption{Frequency with which themes appeared in each interview. }
\label{fig1}
\end{figure}

\begin{figure*}
  \includegraphics[width=0.9\textwidth]{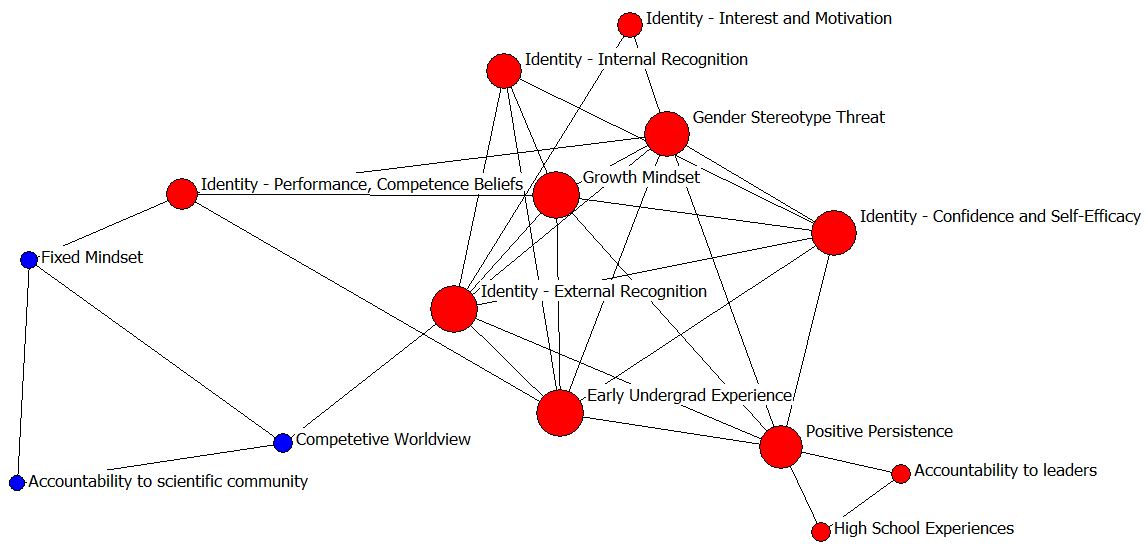}
  \caption{Network map which shows the relationships between codes for female students. \label{fig2}}
\end{figure*}

Notable findings from the interviews with female students are shown in Figures \ref{fig1} and \ref{fig2}. The frequencies with which themes appeared in interviews, with codes counted once per interview, are shown in Figure \ref{fig1}. The semantic network map, at the $p<0.05$ level, is shown in Figure \ref{fig2}. The semantic network map shows growth mindset to be a central theme and is connected to a majority of other nodes in the map. All female students reported characteristics indicative of a growth mindset in regards to challenges they faced or important experiences they had. Students reported that their involvement in informal physics programs helped them to ``demystify science'' and realize that ``anyone can do it'', which can then aid in the formation of a growth mindset, or the idea that abilities can be developed through dedication and hard work \cite{dweck2014mindsets}. We can see this mindset in one student who said, ``Anybody can be a physicist; you just have to be willing to put in the work to learn the really hard stuff.'' Another student expanded on this sentiment by describing their outlook on scientific ability: ``I try to see it as this person is good at this aspect of physics but they're not so good at this other aspect of physics. And if I explain it to them, then they'll get better at that aspect...  it's just a matter of finding the right words to explain it.'' 

Growth mindset was connected to persistence in physics, performance and competency beliefs, and confidence and self-efficacy (Fig. \ref{fig2}). One student shared her experience, saying, ``I actually went into physics because I thought I was really bad at physics... So I just practiced and practiced and took more and more classes until I got better.'' This demonstrates the belief that one's abilities can improve through hard work may help students feel more confident in their abilities. Another student discussing her experiences networking with faculty through outreach said, ``I got to interact with a lot of professors which I think really changed my mind of what a physics person is... realizing they're just normal, very smart people... Okay. I'm normal. Now, all I need to do is get very smart and I can do that through the classes that I'm in.''

The majority of female students reported a positive impact on their confidence and self-efficacy (Fig. \ref{fig1}). One student said about her participation in informal physics programs: ``it made me cement my understanding of, `Okay, this is who I am. I am more confident now. This is how I can present myself.' '' A statement from another student suggests that an increase in confidence can potentially lead to a feeling of empowerment: ``I've definitely been really reaffirmed that it's something that I want to do and something that I can do, something kind of I'm actually able to do.'' 

Most female students also reported on both internal and external recognition (Fig. \ref{fig1}). They discussed external recognition coming from their peers, faculty, and the general public for their roles in informal physics programs. As one student put it, ``[informal physics programs] make you feel more welcomed into the physics department and into the physics community in general. And you just feel like you have a place that you can belong in. You're contributing something. And I think when you feel like you're contributing something to a community, you feel more a part of it.'' Recognition from peers, faculty, and knowing that there are other people like them in the field appears to contribute to physics identity development in terms of internal recognition. One student discussed her experience, saying ``I will say that I met a lot of friends through physics outreach. And a lot of them were girls in physics. And it was kind of cool to meet a lot of people who were having the same thoughts as me, and we could just kind of band together and have our own little community within the physics department.''

The majority of female students reported an increase in their performance and competency beliefs (Fig. \ref{fig1}). Communicating various physics concepts to the general public and applying their skills to real-world applications through the construction and presentation of physics demonstrations may have, according to these students, helped them develop a ``strong grasp of the material''  and ``solidify some of the learning learned in a classroom [in] the real-world setting.'' One student spoke of her belief that the communication skills she gained through facilitation of outreach programs directly contributed to multiple job offers she received at the end of her graduate studies. Through their experiences with informal physics programs, these students came to see themselves as more expert in the field. From one student's experience, ``[Informal physics programs have] really made me feel like I can be a part of the physics major... I think going out and teaching other people physics made me feel like I knew what I was doing, and made me feel like I could keep going on the route of being a physics major.''  

All of the female students reported increases in interest and motivation (Fig. \ref{fig1}), as participation in informal physics programs allowed the students to gain a deeper appreciation of physics and develop their interests. As one student put it, ``You get more involved in more projects, and that only strengthened my love for physics.''

Several female students discussed gender stereotype threat in ways that are best summarized by one student who shared, ``I think the problem with most people in physics, or at least most women in physics, is that they don't really believe that they are worth being there.'' Interviewees who expressed this notion went on to say that their perspectives were changed through their experiences in outreach when they were able to ``embrace curiosity'' and ``meet a lot of people who were having the same thoughts.'' Another student spoke of her experience working with a young girl from an area where girls are underrepresented in science. She spoke of the ``uplifting" and ``empowering" experience of outreach where ``you can make a difference just by being there and expressing your excitement." These types of interactions ``helped me as a woman in science."


From one student's experience, participating in informal physics programs helped to anchor her to the physics major because ``if I hadn't gone to that first outreach event, I would not be where I am right now''. Not only did outreach draw her into the major, but it also helped her to be more ``invested in learning the material'' and made ``learning a lot more fun''. These impacts were seen to center on outreach as an ``outlet'' where she was able to recognize what she had learned in class and observe her improved understanding over time. Overall, outreach helped her to ``do the best I can in college.''

Persistence in physics was linked to important early experiences, both in high school and in the first two years of undergraduate studies (Fig. \ref{fig2}). One student spoke strongly about the influence of her high school physics teacher who gave her a ``pull you aside talk'' and spoke of her talent in the field, encouraging her to ``keep going at it''. This event was a significant influence on her choice to pursue physics as a major in college. Another student discussed her fear and uncertainty in her freshman year that physics ``wasn't the right major'' for her and that she didn't fit ``the mold of the physics major.'' She went on to say that participating in informal physics programs throughout her early undergraduate helped her to realize that ``there's not a specific mold of physics major'' and to feel like she could ``keep going the route of being a physics major''.   

Also connected to persistence in physics was accountability to leaders (Fig. \ref{fig2}). One student said of her experiences, ``I started working with [Dr. X] and that's when I really felt like part of the community, and that's when I started to think, `Oh, I want to continue doing this.'” Persistence was also related to other key ideas such as external recognition and confidence and self-efficacy.

The secondary (blue) cluster contains three codes: accountability to the scientific community, fixed mindset, and competitive worldview (Fig. \ref{fig2}). Related to accountability, several female students discussed a sense of duty they felt to inspire and bring scientific knowledge to a diverse public audience. One in particular spoke of feeling that she ``owe[d] it to people to explain" science as taxes paid for her education and current position. Another student, discussing this cluster of ideas in combination, spoke of her dislike of the competition and fixed mindset found in physics. She talked about her frustration surrounding the portrayal of physicists as ``geniuses'' in media and how that led to her having misconceptions of ``what a physicist is'' and her ``constantly [feeling] stupider than the other students''. She further says, ``you start out in a physics degree thinking that you're brilliant, smarter than everyone else or dumber than everyone else and having to hide it by putting on a lot of bravado.''  She goes on to discuss her ``change in identity to one of a bit more humility'' after realizing through her professors that ``you don't understand everything, but that's okay. None of us do.'' It should be noted that these ideas were discussed not in the context of outreach, but rather in the context of the culture of physics as a whole from the student's experience. It is interesting to note that these two themes of fixed mindset and competitive worldview intersect with the main cluster through the identity aspects of external recognition and performance and competency beliefs.

\section{Discussion}

Through our theoretical framework, discussed in Section II, which consisted of physics identity (recognition, interest and motivation, and performance and competence beliefs), persistence, mindset, worldview, important experiences, accountability, and gender stereotype threat, we explored the impact of facilitation of informal physics programs on women in physics. By applying a semantic network analysis, we observed these elements to be interconnected in specific patterns, as shown in Figure \ref{fig2}.

Our findings suggest that participation in informal physics programs may positively impact all of these areas. Many female students who facilitated informal physics programs self-reported increases in both internal and external recognition since these programs can provide students with opportunities to engage with others in the scientific community and the general public. These interactions led to the students feeling recognized externally and reinforced their own identity as a physics person. Students also had the opportunity to dive into various physics concepts as they applied their knowledge and skills to build and present physics demonstrations, which in turn potentially fostered a deeper understanding of physics and could have led to the increase in students' performance and competency beliefs. The students' experiences cultivating their knowledge in physics may have also led to more appreciation of physics content and applications, which could potentially increase interest and motivation. Within Hazari's framework, this provides the foundation for the continual development of one's physics identity, and prior research shows that a strong physics identity may help women's advancement, persistence, and engagement in physics \cite{kalender2020damage, Kalender2019, hazari2017importance, Hazari2010}.

Physics outreach provides students with the opportunity to collaborate with peers and leaders to share knowledge with the general public, which might have contributed to the majority of students exhibiting a cooperative worldview, or the belief that success comes from working with others for a common benefit. Prior research has demonstrated how the competitive culture of physics may contribute to women feeling unable to fully participate in the field \cite{Lewis2016, barthelemy2016}. A cooperative worldview may help students to avoid comparing themselves to others and to feel less threatened in their position within physics and could be important in creating a more welcoming and inclusive culture.

Our findings showed that recognition, confidence, and self-efficacy are interdependent, as being seen as a physics person may help students feel more confident in their physics abilities. This parallels previous results from Kalender et al., who reported that recognition from influential people such as professors and teaching assistants correlates to physics self-efficacy for female students \cite{Kalender2019}. Students reported that experiences facilitating informal physics programs reaffirmed their goals and skills, which may have improved their confidence and self-efficacy. Prior research has shown that students' physics self-efficacy is closely related to their sense of belonging and impacts their physics identity \cite{Kalender2019}. As indicated by Sawtelle and colleagues, increased physics self-efficacy may also lead to success in the classroom and retention in introductory physics \cite{Sawtelle2012}. 

The facilitation of informal physics programs grants students the opportunity to develop and curate their own sense of expertise as they solidify their understanding to communicate ideas and teach physics to others. These experiences, which allow students to continually learn, grow, and challenge themselves as physicists, could also assist in the development of a growth mindset towards intelligence and scientific ability. Additionally, internal and external recognition gained through participation in informal physics programs may have helped students change their perspective on what it means to be a physicist. Many students discussed their belief before participating in outreach that physics majors looked and acted a certain way, or fit into a ``mold,'' and that physics wasn't the right fit for them. Connected to this is our finding that after participating in outreach, they realized that there is no one mold of a physics major and that physics can be for them. This changing perspective of what it means to be a physics major may in part be due to experiences in outreach encouraging the continual development of a growth mindset toward scientific ability, and may also be important when considering women's sense of belonging in physics. 

As discussed by Lewis and colleagues, stereotypes about scientists can lead to women feeling incompatible with the domain, unable to participate without compromising who they are or want to be, which can then reduce their sense of belonging \cite{Lewis2016}. In our study, students discussed how after participating in informal physics programs, they felt more compatible with the physics major and that they had a place where they belonged. The opportunity for women provided by outreach to engage with other people who share similar experiences within the field of physics and to interact with role models such as professors and teaching assistants may bolster their sense of belonging in physics. As noted by Maries and colleagues, this process of developing a woman's growth mindset, in addition to improving her self-efficacy and increasing her sense of belonging, has the potential to fight gender stereotype threat \cite{Maries}. 

Our findings showed the majority of students reported a positive impact of participation in informal physics programs on their persistence in physics. This is an important finding, as sense of belonging in physics has also been found to be linked to persistence \cite{Lewis2016, lewis2017fitting}. We found that persistence was interconnected with important early experiences, recognition, and accountability to leaders---recognition and support from leaders such as professors, teaching assistants, and high school teachers may have encouraged female students to persist in the field. This parallels results from Avraamidou, whose findings indicated that recognition can come from important early life experiences with teachers, family, social community, etc. \cite{avraamidou2021identities}, and also mirrors results from Hazari et al., who reported on the importance of recognition from high school physics teachers for female students' persistence and identity in physics \cite{hazari2017importance}. We found that persistence was additionally interdependent with a growth mindset. The findings discussed here may have important implications for physics educators and potentially open new areas of research on the interdependent nature of multiple elements of physics identity development and persistence among women in physics.


\section{Conclusion}
A brief survey of physics departments, laboratories, and conferences around the country will quickly reveal that women are severely underrepresented in the field. While the causes and implications of this continuing imbalance have been the focus of much attention both in and out of research literature \cite{NSF, hodapp2015women}, it remains a complex problem to examine. As noted earlier in the paper, female students are subject to gender stereotype threats, can lack sufficient role models or encouragement to pursue and persist in physics, and have been observed to possess lower physics self-efficacy even when they are performing well in their courses. The impact of these disadvantages can be seen in female students' physics identity, their sense of belonging, and their desire to persist in the field. Recent studies have shown that students who facilitate informal physics programs can experience positive gains to their physics identity \cite{Fracchiolla2020, rethman2021impact, randolph2021, perry2021}. Our previous work \cite{rethman2021impact} suggested that these gains may particularly address challenges faced by female physics students. In this study, we narrowed our focus to examine the relation between facilitating physics outreach programs and female student physics identity, persistence, and mindset.

Our mixed-method study showed that female students who facilitated informal physics programs experienced a statistically signiﬁcant shift in their conﬁdence of choice of major after participating in the programs. All female students who were interviewed discussed positive effects of participating in informal programs on their interest and motivation with regards to the ﬁeld of physics and the development of characteristics indicative of the growth mindset. The majority of female students reported a positive impact on their conﬁdence and self-efﬁcacy, internal and external recognition, persistence, performance, and competency beliefs. Furthermore, we found that all of these aspects were interdependent. These interdependencies merit further study in different contexts with a broader range of students to develop a better understanding of the connections between constructs including physics identity and persistence with student facilitation of informal physics programs. 

While this study provides encouraging insight into the impact of facilitation of informal physics programs on female students, there are limitations that should be noted. Students were volunteers for at least one informal physics program at a large, land-grant university and self-reported on their beliefs, perceptions, identity, and experiences. Demographic information beyond gender and classification (undergraduate or graduate) such as ethnicity, first-generation status, etc. was not collected. Lastly, it should be noted that from the principles of social science research, generalizability is not a goal of analysis of qualitative data based on interviews.

There remain rich avenues for future research on the impacts of informal physics programs. In particular, we see a need for studies that can draw out the experiences of other underrepresented groups in physics. A broad goal of this work could be to encourage physics departments to maximize the impact of their informal programs, which are less structured and do not require significant changes to departmental curricula to implement. By intentionally leveraging informal experiences, departments may enhance the identity development and persistence of female students, leading to a more diverse discipline.

\begin{acknowledgments}
This work was supported in part by the Texas A\&M University College of Science. We would like to thank all female student volunteers who participated in this study. We would like to acknowledge the contributions of Daniel Choi and Emily Hay for assisting with conducting and coding the interviews. 

\end{acknowledgments}

\bibliography{bibliography}

\end{document}